\documentclass[conference]{IEEEtran}
\IEEEoverridecommandlockouts
\usepackage{cite}
\usepackage[utf8]{inputenc}
\usepackage{amsmath,amssymb,amsfonts}
\usepackage{amsmath}
\usepackage{graphicx}
\usepackage{textcomp}
\usepackage{xcolor}
\usepackage{booktabs}
\usepackage{optidef}
\usepackage{algorithm}
\usepackage{algpseudocode}
\usepackage{lipsum}
\usepackage[top=0.75in, bottom=1in, left=0.62in, right=0.62in]{geometry}
\def\BibTeX{{\rm B\kern-.05em{\sc i\kern-.025em b}\kern-.08em
    T\kern-.1667em\lower.7ex\hbox{E}\kern-.125emX}}
\begin{document}

\title{HAPS-Enabled Sustainability Provision in Cellular Networks through Cell-Switching}

\author{\IEEEauthorblockN{Görkem Berkay Koç$^{1,2}$, Berk Çiloğlu$^{1,2}$, Metin Öztürk$^1$, Halim Yanikomeroglu$^2$}
\fontsize{10}{10}\selectfont
\IEEEauthorblockA{$^1$Electrical and Electronics Engineering, Ankara Yıldırım Beyazıt University, Ankara, Türkiye\\
\IEEEauthorblockA{$^2$Non-Terrestrial Networks (NTN) Lab, Systems and Computer Engineering, Carleton University, Ottawa, Canada}}
}

\maketitle
\begin{abstract}
There is a consensus in the literature that cell-switching is a viable solution to tackle the draconian increase in the energy consumption of cellular networks.
Although the literature is full of works addressing the energy consumption problem via cell-switching, where small cells with low or no load are turned off and the traffic is offloaded to either adjacent base stations or macro cells, in terrestrial communication networks, they may not be feasible when there is either a lack of infrastructure or insufficient capacity due to high demand. 
The integration of non-terrestrial networks (NTN) into the cell-switching process can be considered as a visionary approach to handle this problem. 
In this regard, high altitude platform station (HAPS) draws considerable attention with its massive footprint, high capacity, and ubiquitous connectivity. 
The aim of this study is to show the potential benefits of using HAPS in cell-switching methods by being a bountiful host for offloaded users from cell-switching operations. More specifically, HAPS is included in the network so that it can increase the switching off opportunities by providing extra coverage and capacity.
The simulation results demonstrate that a significant amount of reduction in energy consumption (as high as 16\%) is obtained while ensuring quality-of-service (QoS) requirements.
\end{abstract}

\begin{IEEEkeywords}
HAPS, sustainability, non-terrestrial networks, 6G, cell-switching, green cellular network
\end{IEEEkeywords}

\section{Introduction}
As a consequence of the rising number of connected devices (e.g., Internet of things (IoT), machine-type communications, etc.), radio access networks are failing to keep pace with the growing demand~\cite{IoTdemandtraffic}. 
To address this issue, mobile network operators constantly look for opportunities for capacity enhancement.
In the fifth generation (5G) of cellular communication networks, one of the most important approaches to increase network capacity is to add new infrastructures such as deploying a tremendous number of small cells (SCs) into the network---called \textit{network densification}~\cite{5Gnewsmallcells}. 
This, in turn, increases the energy consumption of the network, resulting in an escalation in the use of fossil fuels that poses a serious threat to sustainability efforts.

Climate change is a major crisis threatening the survival of life on Earth, and thus many organizations and states aim to zero their carbon emissions before the 2050s~\cite{climatechange}. 
In this context, the telecommunication industry is also trying to reduce its surging energy consumption due to the aforementioned network densification by various methods, including traffic offloading and cell-switching~\cite{telecomindustry}. 
It is important to note that even these techniques cannot offer a sustainable solution in terms of providing energy efficiency and fulfilling the capacity necessities.
Therefore, there is a need for new infrastructures armed with renewable energy sources and new technologies/solutions that can meet the varying traffic requirements.

At this point, uncharted potential benefits of non-terrestrial networks (NTN) may rise to the surface due to their renewable energy-based nature, huge coverage, and ubiquitous connectivity. 
They are not only eco-friendly networks but also a remedy for coverage and connectivity issues that terrestrial networks (TN) struggle with.
As a member of the NTN division, high altitude platform station (HAPS) has drawn considerable attention from both industry and academia, with cutting-edge developments in telecommunication and hardware technologies that made it possible~\cite{main_survey}.
HAPS is a stratospheric aircraft that serves at an altitude of around 20 km from the Earth ground with a quasi-stationary position by promising a ubiquitous connectivity as a super macro base station (SMBS)~\cite{main_survey}.
In addition to being used as SMBS for communication purposes, HAPS can be used as portable data centers, machine learning platforms, etc.
HAPS is considered to be a complementary network component that integrates with TN using its gargantuan footprint and capacity\footnote{The footprint of HAPS varies between 40 km to 100 km for high throughput, yet it can rise to 500 km according to the International Telecommunications Union (ITU) \cite{main_survey}.}.
Since HAPS is envisioned to maintain its service for several months or years after taking off, the energy source for HAPS becomes a key aspect~\cite{main_survey}.
Although a diverse set of energy sources are on the table for HAPS, solar-sourced power is considered a major energy source when the wide surface area of HAPS, is taken into account~\cite{solar_panel}.

The works in the literature have focused on reducing energy consumption caused by base stations (BSs) with a special emphasis on cell-switching methods~\cite{ultra_dens,CS_Last,CS_2011}. The main principle in the cell-switching strategies is that idle or lightly loaded cells are switched off by leveraging their traffic to the available adjacent neighbor cell(s). 
In a more visionary manner, the cell-switching techniques can be applied in vertical heterogeneous networks (VHetNet), given NTN's flexible deployment, coverage, and capacity advantages. 
In conventional TN, offloading may become a challenging issue in various cases of network conditions in terms of density. 
On the one hand, since there are not many BS deployments in rural areas, covering idle users\footnote{The users that are normally served by switched-off BSs are referred to as idle users hereafter.} is not always possible.
On the other hand, although it seems like there is an abundant opportunity of offloading the traffic of deactive (switched-off) BSs in ultra-dense networks, there may not be enough capacity at BSs to accommodate the idle users due to their own already-high data traffic. 

In this work, we argue that HAPS can be a solution to this challenge with its high capacity and coverage, such that the idle users can be offloaded to HAPS when necessary.
In this regard, HAPS enables the switching off\footnote{In this paper, switching off refers to putting BSs into the sleep mode.} the BSs that should normally be kept on with the conventional TN cell-switching methods (due to the above-mentioned reasons), providing less carbon emission.
More specifically, we employ a conventional way of cell-switching approach in a TN, however,---as the main contribution and novelty of this work---HAPS is utilized as another capacity provider to idle users, which, in turn, boosts the cell-switching opportunities. In other words, cell-switching is applied in VHetNet.
Moreover, since HAPS utilizes clean energy sources, the magnitude of saving on carbon emission is multiplied. 

\subsection{Related Works}
Although the literature about HAPS and sustainability is immature, there are some important works available. 
In \cite{cihan}, one of the studies on the field of HAPS in sustainability, the authors studied the potential of HAPS-SMBS to accommodate the unexpected traffic demands of users in TN. Hence, the authors demonstrated the contribution of HAPS to the sustainability of the network through economical, environmental, and social analyses. 
The authors in \cite{canhaps6g} presented how the presence of HAPS could affect sustainability in the environment of traditional micro and macro BSs by using a real-life data set of data traffics.
The study in \cite{intaerial} examined the capability of HAPS offloading in lowering grid energy demand and mobile network operating costs by taking advantage of their renewable energy use.

Unlike HAPS for sustainability in cellular networks, there is an avalanche of works about the cell-switching concept for TN in the literature.
In \cite{ultra_dens}, authors studied a reinforcement learning based method to reduce energy consumption by turning off SCs. 
The study in \cite{CS_Last} offered a solution that covers various sleep depth levels to optimize energy efficiency for increasing number of SC deployments.

To the best of our knowledge, the cell-switching method using HAPS-SMBS has not been precisely investigated in the literature.
The novelty of our paper lies in the hypothesis that HAPS-SMBS enables the cell-switching approach, especially in the scenarios of rural and ultra-dense networks where traffic offloading is not always possible. 
In the former, it is unlikely that there exists another BS in a close proximity to cover the idle users, while in the latter, even though there are numerous options to offload the idle users (as the network is dense and there are many BSs within the same vicinity) because the network is ultra-dense all the BSs are almost fully loaded.
Hence, using HAPS integration to TNs can provide promising advantages in terms of cell-switching.
The contributions of this study are as follows:
\begin{itemize}
    \item We formulate the cell-switching with HAPS as an optimization problem that minimizes the energy consumption of the network by considering QoS constraints.
     \item We examine the role of HAPS in cell-switching by considering different amounts of HAPS-SMBS capacities.
     \item We analyze the limits of the gain provided by the use of HAPS-SMBS with different user densities. The investigation is held through sparse and dense network concepts.
\end{itemize}
The rest of this paper is structured as follows: the system model is presented in Section II, while Section III introduces the problem formulation.
Section IV focuses on the simulation parameters, numerical results, and discussions.
Finally, Section~V concludes the paper.

\section{System Model}

\subsection{Network Model}
\begin{figure}
    \centering
    \includegraphics[width=1\linewidth]{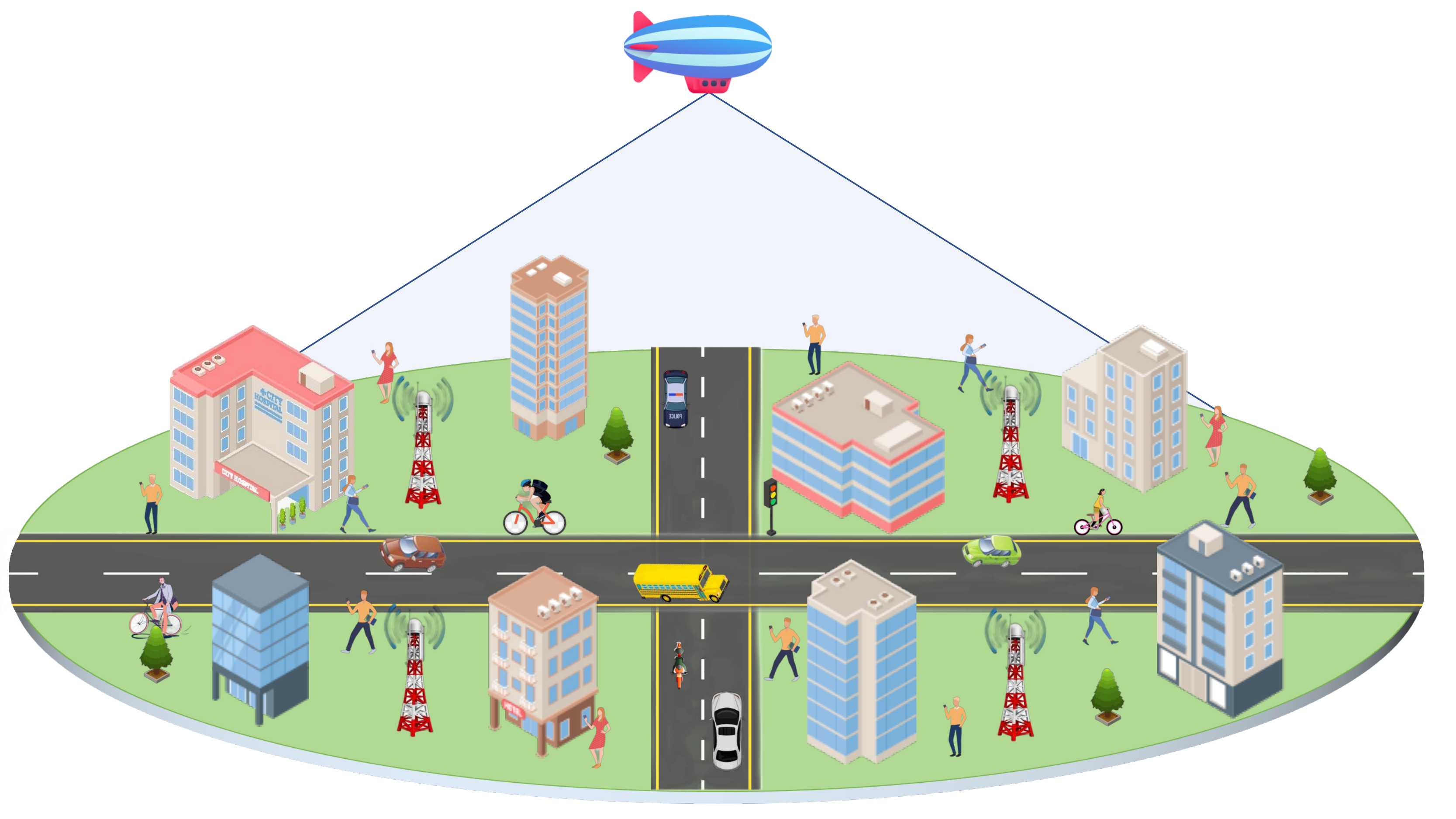}
    \caption{Network model consisting of SCs, HAPS, and users with different mobility modes.}
    \label{fig:networkmodel}
\end{figure}

A two-tier VHetNet consisting of four SCs and a HAPS is considered in this study. SCs are deployed symmetrically with respect to the center of the considered environment.
HAPS is located in the stratosphere by centralizing four SCs (the projection of HAPS on the ground points to the center of the environment).
It should be noted that since HAPS's footprint is huge, it would be unrealistic to assume that the capacity or focus is directed to only our considered environment, and therefore we assign another traffic load to the HAPS that originates from other networks, such that $C_\text{A}=C_\text{T}-C_\text{O}$, where $C_\text{A}$, $C_\text{T}$, and $C_\text{O}$ are the available HAPS capacity to our considered network, total HAPS capacity, and occupied capacity of the HAPS by other networks, respectively.
As the system model is illustrated in Fig.~\ref{fig:networkmodel}, four different user equipment~(UE) mobility modes are considered: stationary, pedestrians, cycler, and vehicular\footnote{Vehicular users are the ones that are traveling in a high-mobility vehicle, such as car, bus, etc.}.

\subsection{Propagation Model}
Since there are two types of networks in the system model; namely, TN and NTN, the path loss models are applied separately.

\subsubsection{Path Loss Model for HAPS (NTN)}
The path loss model for HAPS is taken from 3GPP report given in \cite{3GPP_HAPS}. 
The model is composed of various components and it is heavily dependent on line-of-sight~(LoS) and non-LoS~(NLoS) conditions. 
Thus, it is necessary to have LoS probabilities to evaluate how well HAPS performs in various environments including rural and urban areas.
LoS and NLoS probabilities depend on elevation angles and they can be estimated using the model in~\cite{3GPP_HAPS}.
For simplification, the LoS and NLoS probabilities are defined as $p^\text{LoS}$ and $p^\text{NLoS}$, respectively. Then, the path loss for LoS ($PL^\text{LoS}$) and NLoS ($PL^\text{NLoS}$) conditions can be written as~\cite{linkbudget}
\begin{equation}\label{eq:probPL}
    PL = p^\text{LoS} PL^\text{LoS} + p^\text{NLoS} PL^\text{NLoS}.  
\end{equation}
The signal goes through several stages of propagation and attenuation between HAPS and UE.  
The path loss $PL^\alpha$, $\alpha \in \text{\{LoS,~NLoS\}}$, is shown as~\cite{3GPP_HAPS}
\begin{equation}\label{eq:PL2}
    PL^\alpha = PL^\alpha_\text{b} + PL_\text{g} + PL_\text{s} + PL_\text{e},
\end{equation}
where $PL^\alpha_\text{b}$ is the basic path loss, $ PL_\text{g}$ represents the attenuation caused by the atmospheric gasses,  $PL_\text{s}$ is the attenuation caused by either ionospheric or tropospheric scintillation, and $PL_\text{e}$ denotes the building entry loss.

$PL^\alpha_\text{b}$ depends on the free space path loss ($\operatorname{FSPL}$), clutter loss (CL$^\alpha$), and shadow fading ($X^\alpha$):
\begin{equation}
    PL^\alpha_\text{b} = \operatorname{FSPL} +  \text{CL}^\alpha + X^\alpha, 
\end{equation}
where the $\operatorname{FSPL}$ varies according to the 3D Euclidian distance ($d_\text{3D}$) between HAPS and UE as follows:
\begin{equation}
    d_\text{3D} = \sqrt{R_\text{E}^2 \sin^2{\theta} + H^2 + 2HR_\text{E}} - R_\text{E} \sin{\theta},
\end{equation}
where $R_\text{E}$ is the earth radius, $H$ is the altitude of HAPS, and $\theta$ is the elevation angle. In the LoS condition, CL$^\text{LoS}$ is set to  0 dB, whereas, in the NLoS condition, CL$^\text{NLoS}$ changes by the elevation angle~\cite{3GPP_HAPS}.
$X^\alpha$ is zero-mean Gaussian distribution with standard deviation $\sigma^\alpha$, $\alpha$ $\in \text{\{LoS,~NLoS\}}$, whose values are adopted from~\cite{3GPP_HAPS}. 
Since $PL_\text{g}$ depends on the carrier frequency, according to to~\cite{3GPP_HAPS}, its effect is negligible for carrier frequencies less than 10 GHz (and in this work the sub-6~GHz band is employed, and thus $PL_\text{g}$ is neglected).   
Moreover, because we assume that all the UEs are outdoor, hence $PL_\text{e}$ is neglected as well.

Ionospheric and tropospheric scintillations are the two types of scintillation losses; the former only considerably distorts signals at frequencies below 6 GHz, whereas the latter affects only the signals at frequencies over 6 GHz.
Since 2.5 GHz is used as the carrier frequency in this study, the ionospheric scintillation loss is taken into account as follows:
\begin{equation}
    PL_\text{s} = \frac{\text{PF}}{\sqrt{2}},
\end{equation}
where $\text{PF}$ is peak-to-peak fluctuation, which can be found as
\begin{equation}
    \text{PF}_{(f_\text{c}\le6GHz)} = \text{PF}_{(f_\text{c}=4GHz)}(f_c/4)^{-1.5}, 
\end{equation}
where $f_\text{c}$ is carrier frequency (when $f_\text{c}=4$ GHz, $\text{PF}$ is 1.1 dB~\cite{3GPP_HAPS}).  

\subsubsection{Path Loss Model for TN}
In order to have a fair comparison between the NTN and TN path loss models, the 3GPP report in~\cite{3GPP_Terrestrial} is utilized for TN as well.
According to LoS and NLoS conditions, the path loss of TN is calculated by using~(\ref{eq:probPL}). 
Then, through the link budget calculations, the received power of each UE is given by
\begin{equation}\label{eq:linkbudget}
    P_\text{RX} = P_\text{TX} + G_\text{TX} - PL + G_\text{RX},
\end{equation}
where $P_\text{TX}$ and $G_\text{TX}$ are transmit power and transmitter antenna gain, which are different for HAPS and SCs, respectively, and $G_\text{RX}$ is the antenna gain of UEs. 

\subsection{User Allocation}\label{sec:assoc}
User allocation is performed based on the signal-to-interference-plus-noise ratio (SINR) levels.
In order to associate a user with a BS, there are three conditions to be satisfied: \textit{i)} the BS needs to provide the highest SINR compared to all other BSs; \textit{ii)} the receiver sensitivity requirements of UEs are required to be satisfied; and \textit{iii)} there must be a sufficient available capacity at the BS to be able to accommodate the user.
In this regard, SINR is calculated as a function of $P_\text{RX}$:
\begin{equation}\label{eq:sinr}
    \operatorname{SINR}_{i,j} = \frac{P_\text{RX}^{i,j}}{P_\text{N} + \sum_{k=1,k\neq j}^{n+1} P_\text{RX}^{i,j}},
\end{equation}
where $P_\text{N}$ is additive white Gaussian noise (AWGN) power, $n$ is the total number of SCs, and $i$ and $j$ represent the indices for UEs and BSs, respectively. 
Lastly, the receiver reference sensitivity is considered for the sake of QoS requirements---in order to avoid user outages.

\subsection{Power Consumption Model}\label{powerconsumption}
The EARTH model in~\cite{earth} is adopted in this work to calculate the power consumption of a SC, such that
\begin{equation}\label{eq:pow1}
P_\text{sum} = 
    \begin{cases}
        P_\text{C} + \xi P_\text{TX}, & 0 < P_\text{TX} < P_\text{max}, \\
        P_\text{s}, & P_\text{TX} = 0,
    \end{cases}
\end{equation}
where $P_\text{C}$ is the constant power consumption, $P_\text{s}$ is the power consumption when BS is in sleep mode, $P_\text{TX}$ and $P_\text{max}$ are instantaneous and the maximum transmit power of BSs, respectively. $\xi$ is the slope of the load-dependent power consumption. 
Note that the model in~\eqref{eq:pow1} is generic and can be applied to all different types of SCs with their unique characteristics in terms of $P_\text{C}$, $P_\text{s}$, $P_\text{TX}$, and $P_\text{max}$.

Each SC is assumed to have a certain number of resource blocks (RBs), which is assumed to be the same for all the SCs in this work.
Hence, the power consumption of an SC is given as
\begin{equation}\label{eq:pow2}
    P_{\text{SC}} = P_\text{C}^\text{SC} + \xi^\text{SC} \rho^\text{SC} P_\text{max}^\text{SC},
\end{equation}
where $\rho^\text{SC}=[0,1]$ is the load of the SCs which is calculated as~\cite{earth}
\begin{equation} \label{eq:rho}
    \rho = \frac{\Lambda_\text{M}}{{\Lambda}_\text{T}} = \frac{P_\text{TX}}{P_\text{max}},
\end{equation}
where $\Lambda_\text{T}$ and $\Lambda_\text{M}$ are the total RBs and occupied RBs, respectively. 

To the best of our knowledge, no power consumption model has been studied for HAPS-SMBS in the literature.
Therefore, the macro BS model of~\cite{earth} is adopted for HAPS-SMBS\footnote{It should be noted that the component-based power consumption of HAPS-SMBS, due to the presence of solar panels and location of HAPS, may differ from the model in~\cite{earth}, thus, new studies are needed in this domain.}.
Considering this, the power consumption model for HAPS-SMBS is as follows:
\begin{equation}\label{eq:pow4}
    P_\text{H} = P_\text{C}^\text{H} + \xi^\text{H} \rho^\text{H} P_\text{max}^\text{H},
\end{equation}
where $\rho^\text{H}=[0,1]$ has the same calculation as in \eqref{eq:rho} with different values. HAPS-SMBS and SCs parameters are given in Table \ref{table:simparameter}.

Therefore, the total power consumption of the VHetNet is as follows:
\begin{equation}\label{eq:totalpower}
P_\text{VHetNet} = P_\text{H} + \sum_{k=1}^{n}P_{\text{sum},k}
\end{equation}
where $P_{\text{sum},k}$ indicates the power consumption of the $k^\text{th}$ SC and $k=\{1,2,...,n\}$ represents indices of SCs.
\section{Problem Formulation}

The goal of this study is to obtain the best policy that reduces the energy consumption of the VHetNet. 
A policy $\eta_t = \{\beta_1, \beta_2, \beta_3, ...,\beta_{n+1} \} $ shows which BSs should be active/deactive at a given time $t$.
$\beta_{k,t} \in \{0,1\}$ indicates the active/deactive status (i.e., 1: active and 0: deactive) of the $k^\text{th}$ BS $B_k\in \mathbf{B}=\{B_1, B_2, ..., B_{n+1}\}$ at time $t$, where $\mathbf{B}$ is a vector containing the identities of BSs. 
$B_1$ is the HAPS-SMBS thus $\beta_1$ is always 1 as it is assumed to be always active.

The optimization problem can then be formulated as
\begin{mini}|s|
    {\eta}{\Big[P_\text{H}+\sum_{k=2}^{n+1} (P_{\text{C},k} + \xi_k \rho_k P_{\text{max},k})\beta_{k,t} + P_{\text{s},k}(1-\beta_{k,t}) \Big]} 
    {\label{eq:opt}}{}
    \addConstraint{0\leq}{\rho}{\leq 1,} 
\end{mini}
where the subscript $k$ of a variable indicates the value of that variable for $B_k$.

Let $\mathbf{U}=\{U_1, U_2, ..., U_{\mu}\}$, where $\mu$ is the number of users, be a vector containing the identities of the users.
At each time slot, the users in $\mathbf{U}$ are associated to the BSs in $\mathbf{B}$ according to the policy given in Section~\ref{sec:assoc}, and a user-cell association matrix $\mathbf{A}_{\mu\times (n+1)}$ is created, where $A_{a,b} \in \{0,1\}$ is 1 if user $U_a$ is associated to BS $B_b$, and is 0 otherwise.
Then, $\mathbf\Lambda_\text{M}$ is calculated by $\mathbf{\Lambda_\text{M}}=\sum \text{column}(\mathbf{A})$.
\section{Performance Evaluation}
In order to evaluate the performance of cell-switching with the integration of HAPS, different simulation campaigns are conducted for the cell-switching approach (CSA) and all-active approach (A3) modes.
In CSA, the exhaustive search (ES) algorithm is applied\footnote{The ES algorithm tries all the cell-switching combinations and finds the one causing the least energy consumption while ensuring that all the users are connected.} in order to determine which SCs to switch on/off, whereas all SCs are always kept active in A3.
$\lambda \in [0,1]$ represents the ratio of $C_\text{T}$ for the network in question (i.e., $C_\text{A}$), and it is worth noting that other networks, which use $C_\text{O}$ portion of $C_\text{T}$, are assumed to spatially distributed sufficiently away from the considered network and hence do not cause interference.
For this study, three different $\lambda$ values are used: $\lambda_1=$ 0.7, $\lambda_2=$ 0.5, and $\lambda_3=$ 0.2.
The simulation parameters are given in Table~\ref{table:simparameter}.
\begin{table}[h]
\centering
\caption{Simulation Parameters} \label{table:simparameter}
\resizebox{1\columnwidth}{!}{%
\begin{tabular}{ll}
\textbf{Parameters} & \textbf{Values} \\ \hline
Environment area    & 500 m $\times$ 500 m     \\
Time slot number ($N_\text{TS}$)   & 100             \\
Time slot duration ($t_\text{d}$)  & 1 s             \\
Carrier frequency ($f_\text{c}$)   & 2.5 GHz         \\
Bandwidth  ($W$)         & 50 MHz          \\
Bandwidth per UE    & 200 kHz         \\
SCs transmit power    & 33 dBm   \cite{SCparameter}      \\
HAPS-SMBS transmit power   & 49 dBm   \cite{3GPP_Terrestrial}        \\
SCs antenna gain    & 4 dBi \cite{SCparameter}           \\
HAPS-SMBS antenna gain  & 43.2 dBi   \cite{3GPP_HAPS}      \\
UE antenna gain     & 0 dBi  \cite{3GPP_Terrestrial}          \\
$\sigma^\text{LoS}$ for SCs &  4 dB \cite{3GPP_Terrestrial}  \\
$\sigma^\text{LoS}$ for HAPS-SMBS  & 4 dB \cite{3GPP_HAPS} \\
$\sigma^\text{NLoS}$ for SCs  &  6 dB\cite{3GPP_Terrestrial}  \\
$\sigma^\text{NLoS}$ for HAPS-SMBS  & 6 dB \cite{3GPP_HAPS}  \\
Receiver reference sensitivity  &  -95 dBm  \cite{reference_sensitivity}\\
SCs constant power ($P_\text{C}^\text{SC}$)             & 56 W           \\
HAPS-SMBS constant power ($P_\text{C}^\text{H}$)            & 130 W        \\
SCs slope of load-dependent power ($\xi^\text{SC}$)             & 2.6               \\ 
HAPS-SMBS slope of load-dependent power ($\xi^\text{H}$)             & 4.7                \\ 
SCs maximum transmit power ($P_\text{max}^\text{SC}$)    & 6.3 W \\
HAPS-SMBS maximum transmit power ($P_\text{max}^\text{H}$)    & 20 W \\
Sleep mode power ($P_\text{s}$)    & 39 W \\
\bottomrule
\end{tabular}
}
\end{table}

\subsection{Performance Metrics}
Three performances of CSA and A3 are evaluated using three different metrics: energy consumption, data rate, and gain.

\subsubsection{Energy Consumption}
The energy consumption values are measured by multiplying the power consumption values obtained via \eqref{eq:pow2} and \eqref{eq:pow4} with the time slot duration of the simulations ($t_\text{d}$). 
The total power consumption of VHetNet is then obtained by \eqref{eq:totalpower}. 

\subsubsection{Data Rate}
The data rate is another performance metric that is utilized in order to show the performance of the cell-switching concept because cell-switching is very prone to reduce the user data rates\footnote{Because some BSs are switched off and their users are associated to other BSs that are kept on, and therefore it is expected that the rates of those users (idle users) are violated.}.
It is measured for all users at each time slot using
\begin{equation}\label{eq: rate}
R_{i,j} = W\log_2\left(1 + \operatorname{SINR}_{i,j}\right),
\end{equation}
where $W$ is the bandwidth allocated to the user $i$ by BS $j$.

\subsubsection{Gain}
This metric symbolizes the percentage differences between the energy consumption of CSA ($E_\text{CSA}$) and the energy consumption of A3 ($E_\text{A3}$). It is calculated as follows:
\begin{equation}\label{gain_formula}
G ~(\%) = \frac{E_\text{A3} - E_\text{CSA}}{E_\text{A3}}\times100. 
\end{equation}
\subsection{Result and Discussion}


The network environment is tested on different user densities considered in VHetNet. 
In order to have a convenient analysis, the investigation on the energy consumption and gain metrics are handled with respect to $\mu/\text{m}^2$. This ratio is denoted by $\gamma$ that varies according to the number of users ($\mu$) which is set to 100, 500, 700, and 1000 in this work. 
\begin{figure}[h]
    \centering
    \includegraphics[width=1\linewidth]{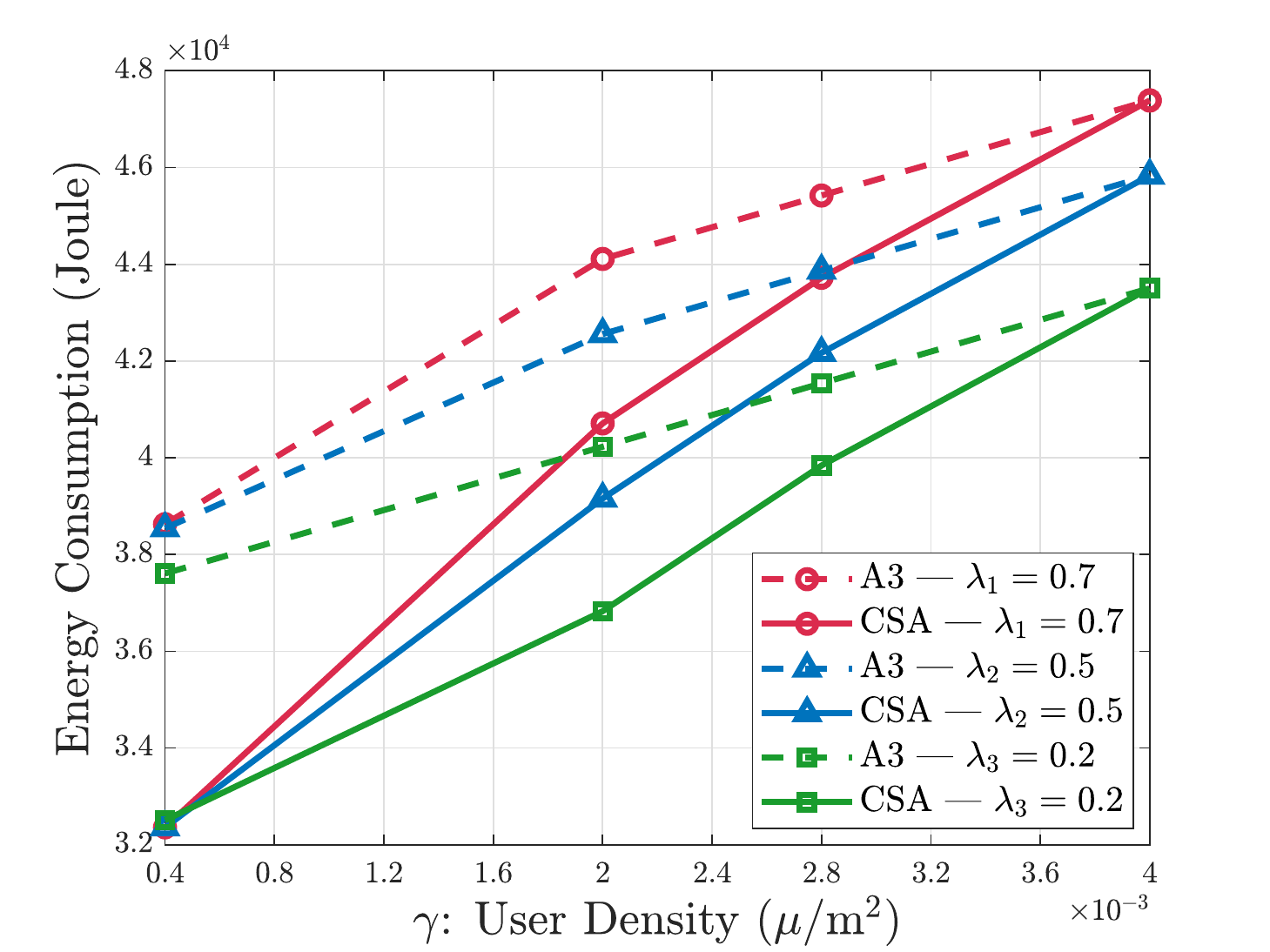}
    \caption{Energy consumption for different approaches and $\lambda$ values.}
    \label{fig:energy}
\end{figure}

Fig.~\ref{fig:energy} demonstrates the amount of dissipated energy with respect to $\gamma$, and includes important takeaways.
First, looking at the results obtained for A3 when $\gamma=0.4\times 10^{-3}$, it is observed that the energy consumption values for $\lambda_1$ and $\lambda_2$ are quite close to each other, but there is a clear difference for $\lambda_3$.
This is because the remaining capacity ($C_\text{A}$) is very low for $\lambda_3$ compared to $\lambda_1$ and $\lambda_2$, and therefore HAPS-SMBS is able to accommodate comparatively less users, resulting in lower power consumption because the load-dependant power consumption of HAPS-SMBS is much higher than that of SCs.
\begin{figure*}[ht!]
    \centering
    \includegraphics[width=1\linewidth]{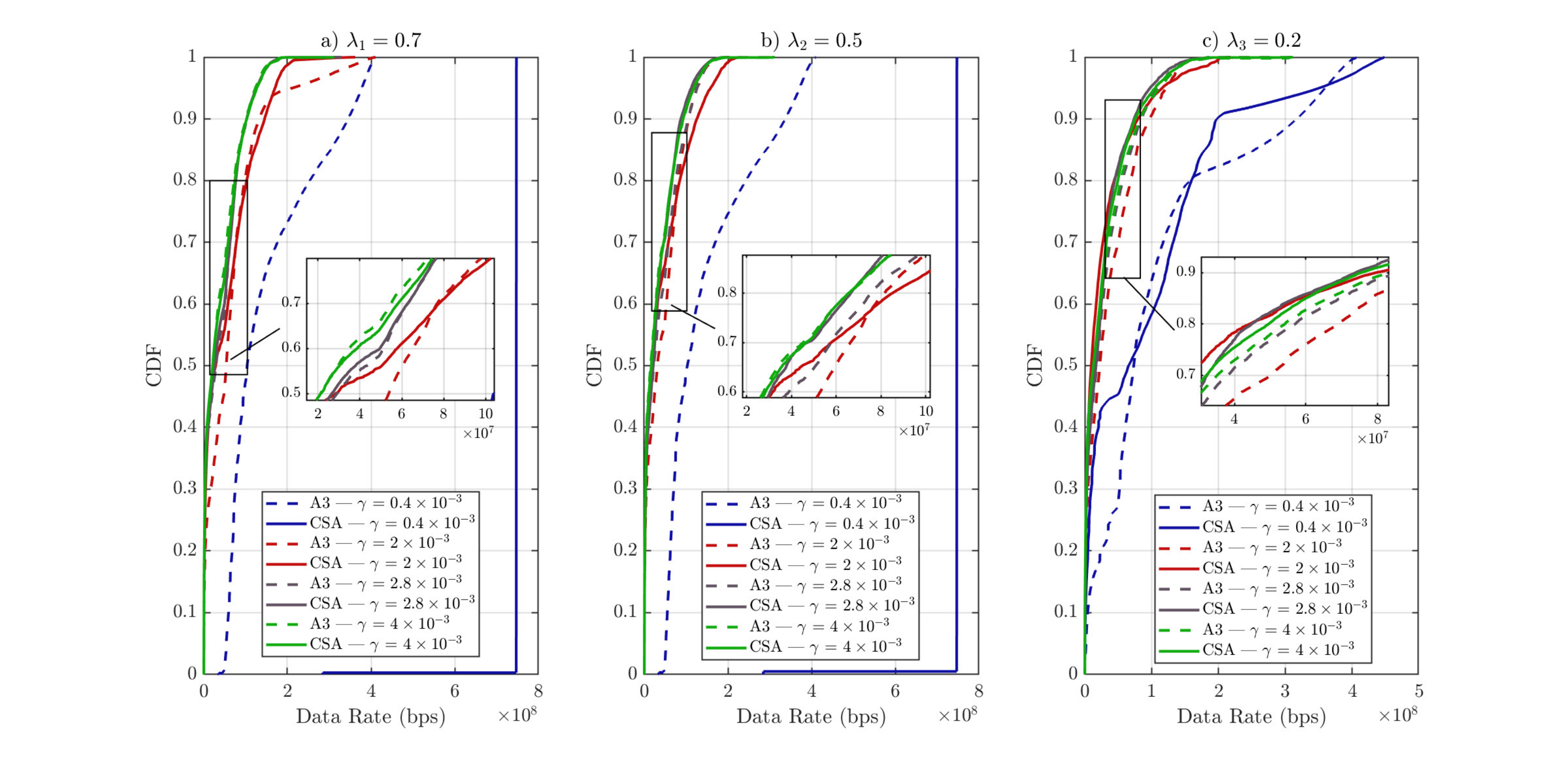}
    \caption{Data rate performance of CSA and A3 for various $\lambda$ values.}
    \label{fig:data_rate}
\end{figure*}
Second, it is obvious from the findings in Fig.~\ref{fig:energy} that on average CSA results in lower energy consumption compared to A3.
However, the difference between the energy consumption of CSA and A3 is more apparent for smaller values of $\gamma$ (i.e., the energy consumption difference fades away as $\gamma$ increases)---the maximum gain obtained when $\gamma=0.4\times 10^{-3}$ is around 16\%.
This originates from the fact that there becomes less switching off opportunities when the number of users increases, and thus more BSs are kept on, which subsequently boosts the overall energy consumption of the network.
More specifically, when $\gamma=0.4\times 10^{-3}$, the average gain is approximately 15\% for all $\lambda$ values, whereas, when $\gamma=4\times 10^{-3}$ value, the average gain drops to 0\%. 
These results reveal that HAPS is a viable companion for cell-switching concept providing significant gain in terms of energy consumption of the network.
Nonetheless, since HAPS-SMBS has limited amount of capacity (i.e., $C_\text{A}$),  cell-switching becomes inapplicable when the network density is high.
This is quite anticipated outcome because when the network density is high, where capacity needs are also high, it does not make sense to switch off BSs at all.

Fig.~\ref{fig:data_rate} presents the data rate differences between CSA and A3 for various $\lambda$ and $\gamma$ values. 
It is worth noting that, for the sake of visualization,  the cumulative distribution function (CDF) of the obtained data rate is divided into three in Fig.~\ref{fig:data_rate} for three $\lambda$ values ($\lambda_1=0.7$, $\lambda_2=0.5$, and $\lambda_3=0.2$). 
Regarding the results for $\gamma=0.4\times 10^{-3}$ in CSA, all the users are associated to HAPS-SMBS in Figs.~\ref{fig:data_rate}a and \ref{fig:data_rate}b, because: \textit{i)} the user density ($\gamma$) is low and \textit{ii)} the amount of available capacity in HAPS-SMBS ($C_\text{A}$) is high.
This brings in more opportunities of switching off SCs and the ES algorithm finds deactivating all SCs an optimum combination; therefore, the dynamic range in user data rates is tight (because all the users are associated to HAPS-SMBS and all SCs are off---no interference).
The reason why a few users receive comparatively less data rates is that they have unfavourable links due to the NLoS situation (remember that the LoS and NLoS situations are determined with respect to a probability in \eqref{eq:probPL}), adversely affecting their data rates considering \eqref{eq:PL2}, \eqref{eq:sinr}, \eqref{eq: rate}.
On the contrary, as seen in Fig.~\ref{fig:data_rate}c, the capacity of HAPS-SMBS is not enough to accommodate all the users, and at least one SC has to be activated. 
This, in turns, causes variations in the data rates---a larger dynamic range---due to interference between the SCs and between the SCs and HAPS-SMBS\footnote{SCs and HAPS-SMBS operate at the same carrier frequency.}.  
The reason why data rates in Fig.~\ref{fig:data_rate} for A3 decrease with the increasing values of $\gamma$ is that user allocation depends on capacity limitations of BSs; hence when $\gamma$ increases two important effects become visible. 
First, there becomes less switching opportunities (i.e., more SCs are on) that subsequently increases the amount of interference experienced by the users---reduces the data rate by~\eqref{eq:sinr} and~\eqref{eq: rate}. 
Second, according to the user association policy given in Section~\ref{sec:assoc}, users are normally associated to the maximum-SINR-providing BSs. However, when the user density is high, the user may not always be able to associate to the maximum-SINR-providing BS as it is likely to be fully-loaded, which, in turn, reduces the achievable rate. 
Lastly, it is observed that CSA results in quite comparable data rate results with A3, which can be deemed as the normal scenario as all the SCs are kept on.

Considering the results of Figs.~\ref{fig:energy} and \ref{fig:data_rate} together, it becomes obvious that cell-switching with the help of HAPS-SMBS not only significantly drops the energy consumption of the network but also respects the QoS constraints by keeping the user data rates close to the A3 case.
This proves the advantages of using HAPS in providing sustainability to cellular networks.



\section{Conclusion}
 In this study, we presented a hypothesis that HAPS-SMBS would enable energy minimization in cellular networks through the cell-switching concept. 
 To compare the results, CSA and A3 cases were considered.
 In order to prove our hypothesis, cell-switching with HAPS should not only reduce the energy consumption of the network but also respect QoS requirements by avoiding significant data rate drops. 
 The results showed that cell-switching with HAPS results in less network energy consumption while not violating data rates.
 Moreover, compared to dense networks (i.e., the user density is high), both energy consumption and data rate performances of cell-switching were better in sparse networks (i.e., the user density is low).
 With the intention of seeing the advantages of HAPS in the cell-switching method, it can be tested in more complex network situations and realistic traffic cases in future studies.
 
\section*{Acknowledgment}
This study is supported in part by the Study in Canada Scholarship (SICS) by Global Affairs Canada (GAC).

\bibliographystyle{IEEEtran}
\bibliography{output}
\end{document}